\documentclass[prd,a4paper,aps,showpacs,floats,superscriptaddress,nofootinbib]{revtex4}
\usepackage{graphicx,  bm, color, amsmath, amssymb, xfrac, calc}
\usepackage[usenames,dvipsnames]{xcolor}

\usepackage{epsfig}
\usepackage{placeins}
\usepackage{amssymb}
\usepackage{color}
\usepackage{float}
\usepackage{hyperref}
\usepackage{yfonts}
\usepackage{amsmath}
\usepackage{MnSymbol}
\usepackage{amssymb}
\usepackage{amsfonts}
\usepackage{eufrak}
\usepackage{bm}
\usepackage{phoenician}
\usepackage{relsize}
\usepackage{hyperref}
\hypersetup{
    colorlinks = true,
    citecolor = {MidnightBlue},
    linkcolor = {BrickRed},
    urlcolor = {BrickRed}
}

\newcommand{\ho}{{H_0}}

\newcommand{\ud}{\mathrm{d}}
\newcommand{\p}{\partial}
\newcommand{\cH}{\mathcal{H}}
\newcommand{\Perp}{\mathcal{P}}

\newcommand{\M}{\mathcal{M}}
\def\VEV#1{\left\langle #1 \right\rangle}

\newcommand{\W}{\mathcal{W}}
\newcommand{\Gtf}{\mathcal{G}}

\newcommand{\D}{\mathcal{D}} 
\newcommand{\Q}{\mathcal{Q}}

\newcommand{\Omo}{\Omega_{m0}}
\newcommand{\OLo}{\Omega_{\Lambda0}}

\DeclareFontFamily{U}{rcjhbltx}{}
\DeclareFontShape{U}{rcjhbltx}{m}{n}{<->rcjhbltx}{}
\DeclareSymbolFont{hebrewletters}{U}{rcjhbltx}{m}{n}

\let\aleph\relax\let\beth\relax
\let\gimel\relax\let\daleth\relax

\DeclareMathSymbol{\aleph}{\mathord}{hebrewletters}{39}
\DeclareMathSymbol{\beth}{\mathord}{hebrewletters}{98}
\DeclareMathSymbol{\gimel}{\mathord}{hebrewletters}{103}
\DeclareMathSymbol{\daleth}{\mathord}{hebrewletters}{100}

\DeclareMathSymbol{\lamed}{\mathord}{hebrewletters}{108}
\DeclareMathSymbol{\mem}{\mathord}{hebrewletters}{109}
\DeclareMathSymbol{\ayin}{\mathord}{hebrewletters}{96}
\DeclareMathSymbol{\tsadi}{\mathord}{hebrewletters}{118}
\DeclareMathSymbol{\qof}{\mathord}{hebrewletters}{114}
\DeclareMathSymbol{\shin}{\mathord}{hebrewletters}{152}

\newlength{\depthofsumsign}
\newcommand{\nsum}[1][2.1]{
    \mathop{%
        \raisebox
            {-#1\depthofsumsign+1\depthofsumsign}
            {\scalebox
                {#1}
                {$\displaystyle\sum$}%
            }
    }
}

\def\be{\begin{equation}}
\def\ee{\end{equation}}

\definecolor{darkred}{RGB}{175,0,0}
\definecolor{darkblue}{RGB}{14,0,185}
\definecolor{grey}{RGB}{0,139,139}

\begin{document}


\title{Cosmological perturbation effects on gravitational-wave luminosity distance estimates}

\author{Daniele Bertacca}
\affiliation{Argelander-Institut f\"ur Astronomie, Auf dem H\"ugel 71, D-53121 Bonn, Germany}

\author{Alvise Raccanelli$^\star$\let\thefootnote\relax\footnote{$^\star$Marie Sk\l{}odowska-Curie fellow}
} 
\affiliation{Institut de Ci\`encies del Cosmos (ICCUB),Universitat de Barcelona (IEEC-UB), Mart\'{i} Franqu\`es 1, E08028 Barcelona, Spain }
\affiliation{Department of Physics \& Astronomy, Johns
	Hopkins University, 3400 N.\ Charles St., Baltimore, MD 21218}


\author{Nicola Bartolo}
\affiliation{Dipartimento di Fisica e Astronomia ``G. Galilei'', Universita' degli Studi di Padova, via Marzolo 8, I-35131,
Padova, Italy}
\affiliation{INFN, Sezione di Padova, via Marzolo 8, I-35131, Padova, Italy}
\affiliation{INAF-Osservatorio Astronomico di Padova, Vicolo dell'Osservatorio 5, I-35122 Padova, Italy}

\author{Sabino Matarrese}
\affiliation{Dipartimento di Fisica e Astronomia ``G. Galilei'', Universita' degli Studi di Padova, via Marzolo 8, I-35131,
Padova, Italy}
\affiliation{INFN, Sezione di Padova, via Marzolo 8, I-35131, Padova, Italy}
\affiliation{INAF-Osservatorio Astronomico di Padova, Vicolo dell'Osservatorio 5, I-35122 Padova, Italy}
\affiliation{Gran Sasso Science Institute, INFN, Viale F. Crispi 7, I-67100 L'Aquila, Italy}

\begin{abstract}
Waveforms of gravitational waves provide information about a variety of parameters for the binary system merging.
However, standard calculations have been performed assuming a FLRW universe with no perturbations. In reality this assumption should be dropped: we show that the inclusion of cosmological perturbations translates into corrections to the estimate of astrophysical parameters derived for the merging binary systems.
We compute corrections to the estimate of the luminosity distance due to velocity, volume, lensing and gravitational potential effects.
Our results show that the amplitude of the corrections will be negligible for current instruments, mildly important for experiments like the planned DECIGO, and very important for future ones such as the Big Bang Observer.
\end{abstract}

\date{\today}

\maketitle

\section{Introduction}
\label{sec:intro}
The existence of gravitational waves (GWs), predicted by Einstein in 1916~\cite{Einstein:1916}, was confirmed one century later
when gravitational-waves were observed by the LIGO-VIRGO collaborations~\cite{Abbott:2016blz, Abbott:2016b, Abbott:2016c}; for a short review of the history of GWs, see~\cite{Cervantes-Cota:2016zjc}.
 
The Advanced LIGO instruments detected GWs from the coalescence of binary Black Holes (BHs); this discovery not only represents a confirmation of the structure of Einstein's General Relativity (GR), but also provides new ways to test it, along with other models of gravity, opening up an entirely new window for observational astrophysics.
More generally, it will be possible to address important questions about our Universe in a novel way. 
Measuring  the waveform from a distant coalescing binary provides us, for example, 
with information on the (redshifted) chirp mass of the source and its luminosity distance.

Gravitational wave astronomy provides a novel window to investigate the Universe and can be used to test cosmological (see e.g.~\cite{Schutz:1986gp, Markovic:1993cr, Finn:1995ah, Cutler:2009qv, Taylor:2012db} for the possibility of determining cosmological parameters form GW observations, \cite{Holz:2004xx, Holz:2005df, Jonsson:2006vc, Camera:2013xfa, Oguri:2016dgk, Baker:2016reh, Fan:2016swi} for  GW luminosity distance-redshift relation and gravitationally lensed  GW sirens,  \cite{Yunes:2013dva, Will:2014kxa, Berti:2015itd, TheLIGOScientific:2016src, Collett:2016dey, Yunes:2016jcc, Ben-Dayan:2016iks, Caprini:2016qxs, Bettoni:2016mij, Sawicki:2016klv} for testing general relativity and modified gravity models with GW, \cite{Raccanelli:2016fmc} for using radio galaxy surveys for GW astronomy)
or astrophysical (e.g.~\cite{Kinugawa:2014, Inayoshi:2016, Hartwig:2016, Konoplya:2016pmh, Raccanelli:2016cud, Cholis:2016kqi, Kovetz:2016kpi}) models.
For some reviews on astrophysical and cosmological studies to be performed using gravitational waves, see~\cite{Buonanno:2014aza,  Maggiore:1999vm, Maggiore:1900zz, Kamionkowski:2015yta, Guzzetti:2016mkm, Bartolo:2016}.

Gravitational waves can also be considered as standard sirens for determining the distance-redshift relation~\cite{Holz:2005df}. Therefore, using future ground- and space-based gravitational wave detectors such as the Einstein Telescope (ET,~\cite{Punturo:2010zz}), LISA~\cite{AmaroSeoane:2012km, Seoane:2013qna, Audley:2017}, the DECI-hertz Interferometer Gravitational-wave Observatory (DECIGO,~\cite{Kawamura:2011zz}) and the Big-Bang Observer (BBO,~\cite{Harry:2006fi}), we will be able to measure the cosmological expansion at good precision up to high redshifts.

It is therefore timely to develop precise formalisms to use GW observations to perform astrophysical and cosmological studies, not only by understanding the investigations that will be enabled by GW-Astronomy, but also by including subtle effects, contaminations and degeneracies (see e.g.~\cite{Hirata:2010ba}).
In this paper we take a further step in this direction, and we analyze the effect of cosmological perturbations and inhomogeneities on estimates of the luminosity distance of black hole binary mergers through gravitational waves.
 We apply the ``Cosmic Rulers'' formalism~\cite{Schmidt:2012ne}, we consider {\it the observer frame} as reference system and we derive 
a different expression {\it wrt} \cite{Laguna:2009re}, which is correct
for the effect of large-scale structures on GW waveforms, accounting for lensing, Sachs-Wolfe, integrated Sachs-Wolfe, time delay and volume distortion effects, and evaluate their importance for future GW experiments.
 Finally we are able to connect our results with the initial amplitude of GW signal
 and interpret the numerical simulations of coalescing BH binaries, which produce the templates, using the observed -- rather than the {\it background} -- frame\footnote{ The background frame  is assumed to be homogeneous and without anisotropies.}.

This paper is structured as follows. In Section~\ref{sec:prop} we setup the formalism to calculate effects of perturbations to the propagation of gravitational waves, and compute their amplitude and phase shifts, in general and in the Poisson gauge.
In Section~\ref{sec:results} we numerically evaluate the corrections for different contributions and at different redshifts, and compare them with predictions for future experiments precision.
We discuss our results and conclude in Section~\ref{sec:conclusions}. \\
Throughout the paper we assume the following conventions: units, $c=G = 1$; signature $(-, +, +, +)$; Greek indices run over $0, 1, 2, 3$, and Latin ones over $1, 2, 3$.


\section{Gravitational waves propagation in the perturbed universe}
\label{sec:prop}
Usually, the effect of perturbations on the propagation of gravitational waves has been often neglected so far; however, with the very recent beginning of the so-called gravitational wave astronomy era, it is timely to start developing precise formalisms to investigate the Universe using GWs.

There have been initial attempts to investigate the Integrated-Sachs Wolfe (ISW) effect on GWs from supermassive black hole mergers and in particular its effect on the system's parameter estimation in~\cite{Laguna:2009re}, the ISW of a primordial stochastic background~\cite{Contaldi:2016koz}, and lensing effects~\cite{Dai:2016igl}; it is worth noting that environmental effects can also affect estimates of the luminosity distance~\cite{Barausse:2014tra}. 
Recently, ~\cite{Bonvin:2016qxr} analyzed the effect of local peculiar velocities on the relation luminosity distance-redshift and on the chirp mass estimate for LISA, in particular referring to the possibility of jointly estimating the two with gravitational waves, as originally suggested in~\cite{Seto:2001qf, Nishizawa:2011eq}.

In this context, in this work we generalize these studies and develop a formalism to compute the change in the estimation of luminosity distance due to the presence of cosmological perturbations and inhomogeneities.
We will use the ``Cosmic Rulers''~\cite{Jeong:2011as, Schmidt:2012ne, Jeong:2012nu, Schmidt:2012nw, Bertacca:2014wga, Bertacca:2014hwa} formalism and calculate the correction to the observed luminosity distance due to volume, lensing and ISW-like effects.


\subsection{Cosmic rulers for gravitational waves}
\label{sec:rulers}
We start by assuming Isaacson's shortwave or geometric optics approximation~\cite{Isaacson:1967zz, Isaacson:1968zza}; in this case, the space-time metric can be written as the sum of two parts: $g_{\mu \nu} =  \tilde g_{\mu \nu} +\epsilon  h_{\mu \nu}$,
where $\tilde g_{\mu \nu}$ is usually named ``background metric" and describes both the FRW metric and first-order perturbations, and $h_{\mu \nu}$  is the gravitational wave metric perturbation (where we are using the same notation of~\cite{Laguna:2009re}). 
In the shortwave approximation and neglecting the response of matter background effect to the presence of  $h_{\mu \nu}$, a gravitational wave can be described as
\begin{equation}
\label{GW-1}
\bar h_{\mu \nu} = A_{\mu \nu} \; e^{i \varphi/\epsilon} = e_{\mu \nu} \;{\cal A}\;e^{i \varphi/\epsilon} = e_{\mu \nu} h\,,
\end{equation}
where $\bar h_{\mu \nu}=h_{\mu \nu}-  \tilde g_{\mu \nu} h/2$ and $h$ is the trace of $h_{\mu \nu}$ w.r.t. the background metric $\tilde g_{\mu \nu}$. Here $e_{\mu \nu}$ is a polarization tensor and ${\cal A}$ and $\varphi$ are real functions of retarded time and describe respectively the amplitude and the phase of the GW (see e.g.~\cite{Laguna:2009re}).

From now on it is convenient  to use the comoving metric $\hat g_{\mu \nu}= \tilde g_{\mu \nu} / a^2$, where  $a$ is the scale factor.
Defining  the GW wave-vector $k_\mu =- \nabla_\mu \varphi$, we have $k^\mu k_\mu =
0$, $k^\mu \nabla_ \mu k^\nu = 0$, $k^\mu \nabla_\mu
e_{\alpha \beta} = 0$ (i.e. the polarization tensor is parallel-transported
along null geodesics) and 
\begin{equation}
\label{GW-2}
\frac{d}{d \chi}\ln {\left({\cal A} a \right)}  = - \frac{1}{2} \nabla_\mu k^\mu \,,
\end{equation}
where we have used $d/d \chi \equiv k^\mu \nabla_\mu$.

We define $x^\mu (\chi)$ as the comoving coordinates in the {\it real frame} (or real space), where $\chi$ is the comoving distance, in real-space, from the source to the detector (the observer).
On the other hand, we call {\it Redshift-GW frame} (RGW) the ``cosmic GW laboratory'' where we perform the observations, i.e. the observed frame\footnote{In the case of photons, this prescription has been used for the first time in~\cite{Jeong:2011as}.}.
In RGW space we use coordinates which effectively flatten our past gravitational wave-cone so that the GW geodesic from an observed BH has the following conformal space-time coordinates:
\begin{equation}
\label{Obframe}
\bar{x}^\mu=(\bar \eta,\; \bar {\bf x})=(\eta_0-\bar \chi, \; \bar \chi \, {\bf n}).
\end{equation}
Here $\eta_0$ is the conformal time at observation, $\bar \chi(z)$ is the comoving distance to the observed redshift in RGW-space, ${\bf n}$ is the observed direction of arrival in the sky of the GW, i.e. $n^i=\bar x^i/\bar \chi=\delta^{ij} (\p \bar \chi/\p \bar x^j)$. Using $\bar \chi$ as an affine parameter in the observed frame, the total derivative along the past GW-cone is $\ud / \ud \bar \chi = - \p/ \p \bar \eta + n^i \p/\p \bar x^i$. Below we will use the subscripts  ``{\it e}" and ``{\it o}" to denote the the observer evaluated at the position where the GW is emitted and at the location of the observer where the GWs are received, respectively.

It is also useful to define the parallel and perpendicular projection operators to the observed line-of-sight direction. 
For any spatial vectors and tensors:
\begin{eqnarray}
\label{Projection}
A_{\parallel} = n^{i} n^{j} A_{ij } \;, \quad
B_{\perp}^i =  \Perp^{ij} B_j = B^i -n^i B_{\parallel}\;,  
\end{eqnarray}
where $\Perp^i_{j}= \delta^{i}_j-n^in_j$. The directional derivatives are defined as
\begin{eqnarray}
\label{Projection2}
\bar \p_\parallel = n^i  {\p \over   \p \bar x^i} \;,  \quad \quad \bar \p^2_\parallel  = \bar \p_\parallel \bar \p_\parallel\;,\quad \quad  \bar \p_{\perp i} =  \Perp^j_i \bar \p_j=   {\p \over   \p \bar x^i} -  n_i \bar \p_\parallel\;, \quad \quad   {\p n^j \over   \p \bar x^i}  &=& \frac{1}{\bar \chi}\Perp_i^j
,\quad \quad \frac{\ud}{\ud \bar \chi} \p_{\perp}^i = \bar \p_{\perp}^i  \frac{\ud}{\ud \bar \chi}-\frac{1}{\bar \chi} \p_{\perp}^i
\;, 
\end{eqnarray}
and we have 
\begin{eqnarray}
\label{Projection3}
 {\p B^i \over   \p \bar x^j}  &=& n^i n_j \bar \p_{\parallel} B_{\parallel} + n^i \bar \p_{\perp j} B_{\parallel} +\bar \p_{\perp j} B_{\perp}^i+ n_j \bar \p_{\parallel} B_{\perp}^i +\frac{1}{\bar \chi} \Perp^i_j B_{\parallel}\;,\nonumber \\
\quad \quad  \bar \nabla^2_\perp &=&\bar  \p_{\perp i} \bar \p_\perp^i = \delta^{ij} {\p \over   \p \bar x^i}  {\p \over   \p \bar x^j} - \bar \partial_\parallel^2 - \frac{2}{\bar \chi} \bar \p_\parallel\;.
\end{eqnarray}
Defining $\bar k^\mu$ as the null geodesic vector in the redshift frame at zeroth order,
\begin{equation}
\label{kmu-0}
 \bar{k}^\mu=\frac{\ud  \bar{x}^\mu }{\ud \bar \chi}=\left(-1, \; {\bf n} \right)\;,
\end{equation}
while for the perturbed case we define the physical $k^\mu$ evaluated at $\bar \chi$ in the following way
\begin{equation}
\label{kmu}
 k^\mu(\bar \chi) = \frac{\ud  x^\mu }{\ud \bar \chi}(\bar \chi)= \frac{\ud }{\ud \bar \chi}  \left(\bar{x}^\mu + \delta x^\mu\right)(\bar \chi) = \left(-1+\delta \nu,\; n^i+\delta n^{i}\right)(\bar \chi).
\end{equation}
Now, let us indicate the apparent positions as $\bar x^\mu$, the observed space (or the redshift-space), while the true positions are at $x^\mu$,  real- space, by the displacement $\Delta x^\mu (\bar \chi)$ of the observed position of  coalescing BH binaries. In other words, we can set up a mapping between RGW- and real- space  (the ``physical frame")  in the following way
\begin{eqnarray}
\label{chi}
\chi = \bar \chi+ \delta \chi\;,  \quad x^\mu (\chi) = \bar{x}^\mu (\bar \chi)+ \Delta x^\mu (\bar \chi)\, , \quad \quad  \Delta x^\mu (\bar \chi) = \frac{\ud  \bar{x}^\mu }{\ud \bar \chi} \delta \chi+ \delta x^{\mu} (\bar \chi)= \bar{k}^\mu \delta \chi+ \delta x^{\mu} (\bar \chi) .
\end{eqnarray}

From Eq. (\ref{kmu}), we obtain explicitly 
\begin{equation}
\label{deltaxmu}
\delta x^{0}(\bar \chi)= \int_0^{\bar \chi} \ud \tilde{\chi} \; \delta \nu (\tilde \chi) \, ,
\quad \quad \quad \quad
\delta x^{i}(\bar \chi)= \int_0^{\bar \chi} \ud \tilde{\chi} \; \delta n^{i} (\tilde \chi)\;,
\end{equation}
where we have imposed the boundary conditions at the observer: $\delta x^{0 (n)}_o=0$ and $\delta x^{i (n)}_o=0$.
In real-space the scale factor is 
\begin{equation}\label{a}
a = a[x^0(\chi)]=a(\bar x^0+ \Delta x^0)=\bar a \left( 1 + \cH   \Delta x^{0}\right)\;,
\end{equation}
where $\bar a = a (\bar x^0)$,
$\cH=\bar a'/\bar a$ and prime indicates $\p/\p \bar x^0 = \p/\p \bar \eta$.
Defining\footnote{ 
For simplicity in the main text we set $a_o=1$. However, for the sake of completeness, we should include the perturbation of the scale factor at observation $\delta a_o= a_o- 1$, since we have to assume that the proper time of the observer at observation $\eta_0$ is fixed by the choice of scale factor via the relation $a(\eta_0) = 1$ in the background.
 In this case our definition of $\Delta \ln a_{\rm txt}$, in the main text, becomes  \[\Delta \ln a  \quad \to  \quad \Delta \ln a= \Delta \ln a_{\rm txt} + \delta a_o\;,  \] 
 where $\delta a_o$ is defined in \cite{Schmidt:2012ne}.
}
\begin{equation}
\label{a}
\frac{a}{\bar a} = 1 + \Delta \ln a \;, 
\end{equation}
we have
\begin{eqnarray}
\label{DeltaLna-1}
\Delta \ln a &=&  \cH \,  \Delta x^{0} =  \cH  \left(  - \delta \chi+ \delta x^{0 } \right)\;.
\end{eqnarray}
Here we consider {\it the local wave zone} approximation to define the tetrads at source position\footnote{The observer ``at the emitted position" is within a region with a comoving distance to the source sufficiently large so that the gravitational field is ``weak enough" but still ``local'', i.e. the gravitational wave wavelength is small w.r.t. the comoving distance from the observer $\bar \chi$.}~\cite{Maggiore:1900zz}. If we choose the four-velocity  $u^\mu$ as the time-like basis vector, then 
\begin{equation}
\label{E0mu}
u_\mu=a \, E_{\hat{0} \mu} \, , \quad \quad \quad \quad u^\mu= a^{-1}E_{\hat{0}}^\mu\;,
\end{equation}
where $E^\mu_{\hat{\alpha}}$ is the tetrad in the comoving frame.  
At the background level we have $E_{\hat{0}\mu}^{(0)}=(-1, {\bf 0})$ and, perturbing the tetrad, we obtain
\begin{eqnarray}
\label{E^mu_hata}
E_{\hat{0}\mu}\left(x^\nu(\chi)\right)=E_{\hat{0}\mu}\left( \bar x^\nu(\bar \chi)+\Delta x^\nu\right)= E_{\hat{0}\mu}^{(0)}(\bar \chi)+E_{\hat{0}\mu}^{(1)}(\bar \chi)\;.
\end{eqnarray}
From $k^\mu$, the map from redshift to real-space is given by
\begin{eqnarray}
\label{pertkmu}
k^{\mu}\left(\chi\right)= \frac{\ud x^{\mu}(\chi)}{\ud \chi}=k^{\mu}\left(\bar \chi+\delta \chi \right)&=& k^{\mu(0)}(\bar \chi)+k^{\mu (1)}(\bar \chi)  \;,
\end{eqnarray}
where $k^{\mu(0)}(\bar \chi)=\bar k^\mu$, and at first order
\begin{eqnarray} 
\label{EqGWphase}
\frac{\ud \delta k^\mu (\bar\chi) }{ \ud \bar \chi}+ \delta \hat \Gamma^\mu_{\alpha \beta}(\bar x^\gamma) \bar k^\alpha(\bar\chi)  \bar k^\beta(\bar\chi) =0 \;,
\end{eqnarray}
where $\delta k^\mu (\bar\chi) =k^{\mu (1)}(\bar \chi) $ and, consequently,
\[\delta k^0 (\bar\chi)= \delta \nu \, , \quad \quad \delta k^i (\bar\chi) = \delta n^i\;. \]
The observed redshift (precisely, if we  have its electromagnetic counterpart) is given by
\begin{eqnarray}
(1+z)=\frac{f_e}{f_o} =\frac{a_o}{a(\chi_e)}\frac{(E_{\hat{0}\mu} k^\mu)\big|_{e} }{(E_{\hat{0}\mu} k^\mu)|_o}\;,
\end{eqnarray}
Choosing $a_o = \bar a_o = a(\bar x^0)=1$ and $(E_{\hat{0}\mu} k^\mu)|_o=1$, we have
\begin{eqnarray}
1+z=\frac{E_{\hat{0}\mu} k^\mu}{a}\;.
\end{eqnarray}
From Eq.\ (\ref{a}), $\bar a$ is the scale factor in redshift-space. Then $\bar a = 1/(1+z)$. From Eqs.\ (\ref{a}), (\ref{E^mu_hata}) and (\ref{pertkmu}), we get 
\begin{eqnarray}
\label{Ek-total}
1=\frac{1+ (E_{\hat{0}\mu} k^\mu)^{(1)}}{ 1 + \Delta \ln a }\;.
\end{eqnarray}
Therefore, we can write $\Delta \ln a$ as
\begin{eqnarray}
\label{Ek-1}
\Delta \ln a &=& (E_{\hat{0}\mu} k^\mu)^{(1)} = E_{\hat{0}\mu}^{(1)} k^{\mu (0)}+E_{\hat{0}\mu}^{(0)} k^{\mu (1)}=- E_{\hat{0}0}^{(1)} + n^i E_{\hat{0}i}^{(1)} - \delta \nu\;,
\end{eqnarray}
Using Eqs.\ (\ref{DeltaLna-1}) and (\ref{Ek-1}) we obtain 
\begin{eqnarray}
\label{chi_1}
\delta \chi & = &  \delta x^{0} - \frac{\Delta \ln a }{\cH}= \delta x^{0}- \Delta x^{0} \;.
\end{eqnarray}
In the rest of this Section we will then use this formalism to compute modifications to amplitude and phase of gravitational waves due to perturbations around a FRW metric.


\subsection{Gravitational waves in the observed frame}
\label{sec:obsgw}
We start by calculating perturbations in a general way.
Let us write $h_{\mu \nu}$ defined in Eq. (\ref{GW-1}) in Redshift-GW frame. First of all, we will compute the phase $\varphi$.  At first order
\begin{eqnarray}
\label{GW-3}
k^\mu k_\mu= - k^\mu \nabla_\mu \varphi =- \frac{\ud}{\ud \chi} \varphi(x^\mu(\chi)) =- \frac{\ud \bar \chi}{\ud \chi} \frac{\ud}{\ud \bar \chi} \varphi(\bar x^\mu + \Delta x^\mu)=- \left(1-\frac{\ud \delta \chi}{\ud \bar \chi} \right)\frac{\ud}{\ud \bar \chi} \left[\bar \varphi + \delta \varphi (\bar x^\mu)+ \Delta x_\mu \bar  \nabla^\mu \bar \varphi  \right]=0\;,
\end{eqnarray}
where $\bar \varphi  =\varphi^{(0)}(\bar x^\mu)$. Defining $\bar k_{\mu} =-\bar \nabla_\mu \bar \varphi$,  $\ud \varphi (x^\mu)/\ud \bar \chi= \bar k^{\mu} \bar \nabla_\mu  \varphi (x^\mu)$ and $ \varphi (\bar x^\mu)=\bar \varphi +  \delta \varphi (\bar x^\mu) $, we find
\begin{eqnarray}
\label{GW-4}
 \frac{\ud}{\ud \bar \chi}\delta  \varphi(\bar x^\mu) = \bar k_{\mu} \delta k^\mu (\bar\chi)\;.
\end{eqnarray}
Note that $\ud \varphi (x^\mu)/\ud  \chi=0$ instead of $\ud \varphi (\bar x^\mu)/\ud \bar \chi \neq 0$. 
Finally, perturbing directly $k^\mu=- \hat{g}^{\mu \nu} \nabla_\nu \varphi$ we get
\begin{equation}\label{GW-5}
\delta k^\mu (\bar\chi) = - \bar g^{\mu\nu} \bar \nabla_\nu \delta \varphi (\bar x^\mu)  + \delta x^\alpha \bar \nabla_\alpha \bar k^\mu+ \delta g^{\mu \nu} \bar k_\nu \;,
\end{equation}
where $\hat g_{\mu\nu}=\bar g_{\mu\nu} + \delta g_{\mu \nu}$ and $\bar g_{\mu\nu}$ is the Minkowski metric\footnote{
It is interesting to show that starting from Eq.(\ref{GW-5}) we get
\[  \frac{\ud}{\ud \bar \chi}\delta  \varphi(\bar x^\mu) = - \bar k_\mu \delta k^\mu +\bar k_\mu \delta g^{\mu\nu}\bar k_\nu \;.\] However,  it is easy to see that perturbing $k^\mu \hat{g}_{\mu \nu} k^\mu=0$ we have $$ \bar k_\mu \delta k^\mu = -{1 \over 2}\bar k^\mu \delta g_{\mu\nu}\bar k^\nu $$ and using $\delta g^{\mu\nu}=  - \bar g^{\mu\alpha} \delta g_{\alpha \beta} \bar g^{\nu\beta}$ we are able to recover Eq. (\ref{GW-4}).}.
\vspace{5mm}

Now we turn to the amplitude; starting from  Eq.(\ref{GW-2}), up to linear order, we have
\begin{eqnarray}
\label{eq:obsgw25}
 \left(1-\frac{\ud \delta \chi}{\ud \bar \chi} \right)\frac{\ud}{\ud \bar \chi} \ln \bigg\{{{\cal A}(\bar x^\mu + \Delta x^\mu) \, \bar a\big[1+\Delta \ln a\big]}\bigg\}  = - \frac{1}{2}\left[ \left(\frac{ \p \bar x^\nu}{\p x^\mu } \right) \frac{\p  }{\p \bar x^\nu} \left(\bar k^{\mu}+\delta k^{\mu}\right) + \delta \Gamma_{\mu \nu}^\mu \bar k^{\nu}\right]\;.
\end{eqnarray}
In particular,  following the prescription defined in the previous subsection, we can divide the contributions to Eq.~\eqref{eq:obsgw25} in three parts
\begin{eqnarray}
\frac{\ud}{\ud \bar \chi} \ln \left[{\cal A}\left(\bar x^\mu + \Delta x^\mu\right)\right]&=&\frac{\ud}{\ud \bar \chi} \ln  \bar{\cal A} + \frac{\ud}{\ud \bar \chi} \left[\delta  \ln{\cal A} +   \Delta x^\mu \bar \p_\mu  \ln  \bar{\cal A} \right] =\frac{\ud}{\ud \bar \chi} \ln  \bar{\cal A} +  \frac{\ud}{\ud \bar \chi} \delta  \ln{\cal A} +   \left(\frac{\ud^2}{\ud \bar \chi^2} \ln  \bar{\cal A} \right) \delta \chi\nonumber \\
&& +  \left(\frac{\ud}{\ud \bar \chi} \ln  \bar{\cal A} \right) \left( \frac{\ud}{\ud \bar \chi} \delta \chi \right) + \left( \frac{\ud}{\ud \bar \chi} \bar \p_\mu    \ln  \bar{\cal A}\right)  \delta x^\mu+   \left(\bar \p_\mu    \ln  \bar{\cal A}\right)  \delta k^\mu\;, \nonumber \\\\
\frac{\ud}{\ud \bar \chi} \ln \left[ \bar a\left(1+\Delta \ln a \right)\right] &=& -\mathcal{H} -\mathcal{H}'  \left(- \delta \chi + \delta x^0 \right) + \mathcal{H} \left(- \frac{\ud}{\ud \bar \chi} \delta \chi + \delta k^0 \right) \;, \nonumber \\\\
 \left(\frac{ \p \bar x^\nu}{\p x^\mu } \right)  \frac{\p}{\p \bar x^\nu} \left(\bar k^{\mu}+\delta k^{\mu}\right)  &=&\left( \delta_\mu^\nu -  \bar \p_\mu \Delta x^\nu \right) \frac{\p}{\p \bar x^\nu} \left(\bar k^{\mu}+\delta k^{\mu}\right)  = \frac{2}{\bar \chi} \left(1 + \delta k_\| \right) + \left(\frac{\ud}{\ud \bar \chi}\delta k_\| \right) + \frac{\p}{\p\bar x^0} \left(\delta k^0 +\delta k_\| \right) + \bar \p_{\perp i} \delta k_{\perp}^{i} \nonumber \\
&&- \frac{2}{\bar \chi^2} \left(\delta \chi + \delta x_\| \right) - \frac{1}{\bar \chi} \bar \p_{\perp i} \delta x_{\perp}^i\;,
\end{eqnarray}
where $  \bar \p_{\perp i} \Delta x_{\perp}^i= \bar \p_{\perp i} \delta x_{\perp}^i$ and $\Perp^i_{i}=2$.
Then to the lowest order we have (see also~\cite{Laguna:2009re})
\begin{equation}
 \bar{\cal A}(\bar x^0, \bar \chi) =  \frac{\Q}{\bar a(\bar x^0) \bar \chi}=\frac{\Q (1+z)}{\bar \chi} \;,
 \end{equation}
 where  $\Q$ is constant along the null geodesic.
 Here $\Q$ is determined by the local wave-zone source solution and contains all the physical information on the spiralling binary\footnote{The physical meaning of $\Q$ is not the same as \cite{Laguna:2009re}. Indeed, in this paper, $\Q$ is defined directly in the observed frame $\bar{x}^\mu=(\bar \eta,\; \bar {\bf x})$ and, in \cite{Laguna:2009re}, with a unperturbed background metric.}.
At the receiving location it is given by the same solution evaluated at the retarded time.
With this result, using Eq.\ (\ref{Projection2}) and
 \begin{eqnarray}
  \left( \frac{\ud}{\ud \bar \chi} \bar \p_\mu    \ln  \bar{\cal A}\right)  \delta x^\mu+ \left(\bar \p_\mu    \ln  \bar{\cal A}\right)  \delta k^\mu =
 \mathcal{H}' \delta x^0 +  \frac{2}{\bar \chi^2}  \delta x_\|  - \mathcal{H}  \delta k^0 - \frac{1}{\bar \chi}  \delta k_\| \; , \\
\bar \p_{\perp i} \delta k_{\perp}^{i} =  \bar \p_{\perp i}  \frac{\ud}{\ud \bar \chi} \delta x_{\perp}^{i} = \frac{\ud}{\ud \bar \chi} \bar \p_{\perp i}  \delta x_{\perp}^{i} +  \frac{1}{\bar \chi} \bar \p_{\perp i}   \delta x_{\perp}^{i} \; ,
\end{eqnarray}
we can finally write
\begin{equation}
\label{GW-6}
 \frac{\ud}{\ud \bar \chi} \delta  \ln{\cal A} = - {1 \over 2}\left[  \frac{\p}{\p\bar x^0}\left( \delta k^0 + \delta k_\| \right) +  \frac{\ud}{\ud \bar \chi}  \delta k_\| -2  \frac{\ud}{\ud \bar \chi} \kappa + \delta \Gamma_{\mu \nu}^\mu \bar k^{\nu} \right] \, ,
\end{equation}
where $\kappa$ is the weak lensing convergence term
 \begin{eqnarray}
 \label{kappa}
\kappa=- \frac{1}{2}  \bar \p_{\perp i} \Delta x_{\perp}^{i} \; .
\end{eqnarray}

%

%
The perturbed gravitational waves can then be fully described as
\begin{equation}
h(\eta_e, {\bf x}_e) = {\cal A}(\eta_e, {\bf x}_e)\;e^{i\varphi(\eta_e,  {\bf x}_e)} = \frac{\Q(1+z)}{\bar \chi}(1+\Delta  \ln{\cal A} )e^{i(\bar \varphi+\Delta\varphi)}\,,
\label{eq:hmunu}
\end{equation}
where the perturbations of amplitude and phase are
 \begin{eqnarray}
 \label{DeltacalA}
 \Delta  \ln{\cal A} &=&\delta  \ln{\cal A} + \Delta x^0 \bar \p_0 \ln  \bar {\cal A} + \Delta x_\|  \bar \p_\| \ln  \bar {\cal A}=\delta  \ln{\cal A} - \left(1-\frac{1}{\cH \bar \chi} \right) \Delta \ln a + \frac{T}{\bar \chi} \; , \\
 \label{Deltacalphi}
  \Delta  \varphi &=&  \delta \varphi + T \; .
 \end{eqnarray}
 with $-T= \Delta x^0 + \Delta x_\| = \delta x^0 + \delta x_\| $.
 
\subsection{Perturbations in the Poisson gauge}
\label{Sec:PoissonGauge}
The relations obtained so far are valid in any gauge. We now derive the expressions for the amplitude and phase corrections in the Poisson gauge, so that we will be able to estimate their practical relevance.
The background metric $\tilde g_{\mu \nu}$ in Poisson gauge reads
\begin{eqnarray} 
\label{Poiss-metric}
\ud  s^2 = a(\eta)^2\left[-\left(1 + 2\Phi \right)\ud\eta^2+ \delta_{ij} \left(1 -2\Psi \right)\ud x^i\ud x^j\right] \;,
\end{eqnarray}
where we are neglecting vector and tensor perturbations at first order.

For the geodesic equation we obtain, at linear order,
\begin{eqnarray} 
\label{Poiss-dkmu}
\frac{\ud}{\ud\bar \chi} \left(\delta \nu - 2 \Phi\right) = \Phi' +\Psi' \;, \quad \quad \quad \quad \quad \frac{\ud}{\ud\bar \chi} \left( \delta n^i
 -2 \Psi n^i \right) = - \bar \p^i \left( \Phi  + \Psi \right), 
\end{eqnarray}

To solve Eq.\ (\ref{Poiss-dkmu}) we need the values of $\delta \nu$ and $\delta n^{i (1)}$   today. In this case we need all the components of the tetrads  $E^{\hat \alpha}_\mu$, which are defined through the following relations (see Appendix \ref{Poiss-pert})
\begin{eqnarray}
\label{E}
\hat g^{\mu \nu} E^{\hat \alpha}_\mu E^{\hat \beta}_\nu& = &\eta^{\hat  \alpha \hat \beta}\;, \quad  \quad  \eta_{\hat  \alpha \hat \beta} E^{\hat \alpha}_\mu E^{\hat \beta}_\nu = \hat g_{\mu \nu}\;,  \quad  \quad \hat g^{\mu \nu} E^{\hat \beta}_\nu = E^{\hat \beta \mu}\;,    \quad  \quad \eta_{\hat  \alpha \hat \beta}  E^{\hat \beta}_\nu  =  E_{\hat \beta \nu}  \;,
\end{eqnarray}
where $\eta_{\hat  \alpha \hat \beta}$ is the comoving Minkowski metric \cite{Bertacca:2014wga}.
Using the constraints
\begin{eqnarray}
\label{Poiss-dnude-o}
\delta\nu_o = \Phi_o+v_{\| o}\;, \quad \quad \quad \quad \quad \delta n^{\hat a }_o=-v^{\hat a }_o + n^{\hat a} \Psi_{ \, o}  \;, 
\end{eqnarray}
from Eq.(\ref{Poiss-dkmu}) we obtain at first order
\begin{eqnarray}
\label{Poiss-dnude-1}
\delta\nu&=&- \left (\Phi_o-v_{\| \, o}\right)+ 2 \Phi  + \int_0^{\bar \chi} \ud \tilde \chi \left(\Phi' +  \Psi' \right) = - \left (\Phi_o-v_{\| \, o}\right)+ 2 \Phi - 2I \;, \\
\delta n^{i}&=& -v^{i}_o- n^i \Psi_{\, o}  +2 n^i \Psi - \int_0^{\bar \chi} \ud \tilde \chi\,  \tilde \p^i\left(  \Phi  + \Psi  \right) = n^i \delta n_\|+\delta n_\perp^{i}\;,
\end{eqnarray}
where
\begin{eqnarray}
\label{Poiss-dnue-||perp-1}
\delta n_\| =\Phi_o-v_{\| \, o}-\Phi+ \Psi+2I \;,   \quad \quad \quad \delta n_\perp^{i}=  -v^i_{\perp \, o }+ 2S_{\perp}^i  \;.
\end{eqnarray}
Here
\begin{eqnarray}
I  = -\frac{1}{2} \int_0^{\bar \chi} \ud \tilde \chi\left(\Phi' +  \Psi' \right) \;, \quad \quad \quad \quad S_{\perp}^{i(1)} = -\frac{1}{2} \int_0^{\bar \chi} \ud \tilde \chi \, \tilde\p^i_\perp \left( \Phi  + \Psi \right)\;
\end{eqnarray}
where, $I $ is the integrated Sachs-Wolfe (ISW) term \cite{Sachs:1967er}.
The GW phase in Eq.~\eqref{GW-4} can be obtained with the following relation
\footnote{To obtain Eq.~\eqref{Poiss-du+de} we
used Eqs.\ (\ref{GW-4}), and we note that
\begin{eqnarray}
\label{s-1}
\delta \varphi -\delta \varphi_o &=&- T=\delta x^{0} + \delta x_{\|} =   \int_0^{\bar \chi} \ud \tilde \chi \left(\Phi + \Psi \right) \; , \\
\label{Poiss-dx0-1}
\delta x^{0}&=& -\bar \chi \left (\Phi_o-v_{\| \, o}\right)+ \int_0^{\bar \chi} \ud \tilde \chi \left[ 2 \Phi + \left(\bar \chi-\tilde \chi\right) \left(\Phi'+ \Psi' \right) \right] \; , \\
\label{Poiss-dx||-1}
\delta x_{\|}&=& \bar \chi \left(\Phi_o-v_{\| \, o}\right)-\int_0^{\bar \chi} \ud \tilde \chi \left[ \left(\Phi - \Psi\right) +  \left(\bar \chi-\tilde \chi\right) \left(\Phi' +\Psi' \right) \right] \;,   \\
\label{Poiss-dxperp-1}
\delta x_{\perp}^i&=& - \bar \chi \, v^i_{\perp \, o }-\int_0^{\bar \chi} \ud \tilde \chi \left[ \left(\bar \chi-\tilde \chi\right) \tilde \p^i_\perp \left( \Phi + \Psi \right) \right] \; .
\end{eqnarray}
}
\begin{eqnarray}
\label{Poiss-du+de}
\frac{\ud \delta \varphi}{\ud \bar \chi}=\delta n_\| +  \delta\nu&=& \Phi+ \Psi \;.
\end{eqnarray}
 From Eq. (\ref{chi_1}) we then have
\begin{eqnarray}
\label{Poiss-Deltalna-1}
\Delta \ln a &=& \left (\Phi_o-v_{\| \, o}\right) - \Phi + v_\| + 2I =\left (\Phi_o - v_{\| \, o}\right) - \Phi + v_\| - \int_0^{\bar \chi} \ud \tilde \chi \left(\Phi' + \Psi' \right)\;, \\
\delta \chi  &=&-\left(\bar \chi+\frac{1}{\cH}\right)\left(\Phi_o-v_{\| \, o}\right)+  \frac{1}{\cH}\left(\Phi - v_\| \right)  + \int_0^{\bar \chi} \ud \tilde \chi \left[ 2 \Phi +\left(\bar \chi-\tilde \chi\right) \left(\Phi' + \Psi' \right) \right] -\frac{2}{\cH}I \;.
\end{eqnarray}
We can therefore write explicitly the components of $\Delta x$; from Eqs. (\ref{a}) and (\ref{Ek-1}) we find
\begin{eqnarray}
\label{Poiss-Dx^0-1}
\Delta x^{0}&=&\frac{1}{\cH}\left[\left (\Phi_o-v_{\| \, o}\right) - \Phi+ v_\|+ 2I\right] =  \frac{1}{\cH}\left[\left (\Phi_o-v_{\| \, o}\right) - 
\Phi + v_\| - \int_0^{\bar \chi} \ud \tilde \chi \left(\Phi'+ \Psi' \right) \right]\; ;
\end{eqnarray}
from Eqs.\ (\ref{chi}) and (\ref{s-1}) we have
\begin{eqnarray}
\label{Poiss-Dx||-1}
\Delta x_\| &=&- T - \frac{1}{\cH}\left[\left (\Phi_o-v_{\| \, o}\right) - \Phi + v_\| + 2I \right]\nonumber \\
&=& \int_0^{\bar \chi} \ud \tilde \chi \left(\Phi +\Psi\right)- \frac{1}{\cH}\left[\left (\Phi_o-v_{\| \, o}\right) - \Phi + v_\| - \int_0^{\bar \chi} \ud \tilde \chi \left(\Phi'  + \Psi' \right) \right] \, ,
 \end{eqnarray}
and 
 \begin{eqnarray}
 \label{Deltax_perp}
 \Delta x_{\perp}^i&=& \delta x_{\perp}^i= - \bar \chi \, v^i_{\perp \, o }-\int_0^{\bar \chi} \ud \tilde \chi  \left(\bar \chi-\tilde \chi\right)  \tilde \p^i_\perp \left( \Phi + \Psi \right)  \;.
\end{eqnarray}
We can recognize that in Eq.\ (\ref{Poiss-Dx^0-1}) there is an ISW contribution and in Eq.\ (\ref{Poiss-Dx||-1}) we have both time-delay and ISW contributions, while Eq.\ (\ref{Deltax_perp}) represents the lensing contribution.

Finally, from Eq.\ (\ref{GW-6}), we can see that $\delta  \ln{\cal A} - \delta  \ln{\cal A}_o= \Psi-\Psi_o + \kappa$.
Therefore, we can express perturbations in amplitude and phase of Eqs.~(\ref{DeltacalA}) and (\ref{Deltacalphi}), in the Poisson gauge, as
\begin{eqnarray}
\label{DeltacalA2}
\Delta  \ln{\cal A} &=& \delta  \ln{\cal A}_o +\Psi-\Psi_o + \kappa - \left(1-\frac{1}{\cH \bar \chi} \right) \Delta \ln a + \frac{T}{\bar \chi}\;, \\
\label{Deltacalphi2}
\Delta  \varphi &=&  \delta \varphi_o \; .
\end{eqnarray}
The correction in Eq. (\ref{DeltacalA2}) can be related to the luminosity distance $\D_L$, in the following way
\begin{eqnarray}
\label{deltaDL}
\frac{ \Delta \D_L}{\bar \D_L} = -\Psi -  \kappa +\left(1-\frac{1}{\cH \bar \chi} \right) \Delta \ln a -   \frac{T}{\bar \chi} \, ,
\end{eqnarray}
taking into account that 
\footnote{
Here we have assumed  $ \delta \varphi_o =0$, i.e. $\varphi_e= \bar  \varphi$. Moreover, by construction, we have
$\delta  \ln{\cal A}_o = \Psi_o$.
}
\begin{equation}
\label{h2}
h_e= \frac{\Q(1+z)^2}{ \D_L} e^{i\bar \varphi}\,.
\end{equation}
and 
\begin{equation}
\Delta  \ln{\cal A} =- {\Delta \D_L \over \bar \D_L} \, ,
\end{equation}
defining $\bar \D_L = (1+z) \bar \chi$  the observed  average luminosity distance  taken over all the sources with the same observed redshift  $z$, with $\D_L=\bar \D_L + \Delta \D_L$.

The gravitational wave observed at the detector is red-shifted, hence we find
\begin{equation}
\label{h2}
h_r \equiv \frac{h_e}{(1+z)} = \frac{\Q(1+z)}{ \D_L} e^{i\bar \varphi}\,.
\end{equation}
Now, we have to estimate correctly $\Q$ from inspiral of compact binaries~\cite{Poisson:1995ef}. For simplicity, in this work, {\it i)} we assume the Newtonian approximation which agrees with standard weak-field approximation in general relativity (i.e., we are neglecting the post newtonian terms) and {\it ii)} we consider only the regime called of ``quasi-circular" motion (i.e. the approximation in which a slowly varying orbital radius is applicable) \cite{Thorne:1969rba, Thorne:1987af, Maggiore:1900zz} .
Then from the quadrupole formula we have \cite{Poisson:1995ef}
\begin{equation}
\label{Q}
\Q=\M_e \left( \pi f_e \M_e \right)^{2/3}\, ,
\end{equation}
where ${\cal M}_e$,  and
$f_e$ are the intrinsic ``chirp mass'' and frequency of the binary, respectively.
 Then $\bar \varphi= \varphi_c
-(\pi\,f_e\, {\cal M}_e)^{-5/3}/16$ where $\varphi_c$ is the value of the phase at $f=\infty$ and 
$t(f) = t_c - (5/256) {\cal M}_e (\pi f_e  {\cal M}_e)^{-8/3}$ \cite{Thorne:1987af}.

A note is in order here. The relation in Eq. (\ref{Q}) formally is the same as that of  \cite{Laguna:2009re}, but its physical meaning is totally different. In fact in \cite{Laguna:2009re} the authors consider as redshift the inverse of scale factor in the background frame, i.e. a Universe without  inhomogeneities and anisotropies. Instead Eq. (\ref{Q}) depends on the {\it measured} redshift. In other words, this value of redshift coincides with that of a hypothetical event where it is possible to measure its electromagnetic signal, i.e. the photons from the coalescence.  Finally, from Eq. (\ref{deltaDL}), we immediately  note that we have recovered the same result obtained by the luminosity distance computed for the photon, see Eq. (51) in ref. \cite{Schmidt:2012ne}.

Taking into account that  $\Q$ is computed in the observed frame, defining $\M_r=\M_e(1+z)$ and $f_r=f_e/(1+z)$, we have\footnote{i.e. $f_o=f_r$ and $\bar \varphi =  \varphi_r$}

\begin{equation}
\label{h2}
h_r = \frac{\M_r}{ \D_L} \left( \pi f_r \M_r \right)^{2/3} e^{i \varphi_r}\, ,
\end{equation}
and considering that $\bar h_r = \M_r \left( \pi f_r \M_r \right)^{2/3} e^{i \varphi_r}/\bar \D_L\,$, we have
\begin{equation}
\label{deltah}
\frac{\Delta h_r}{\bar h_r } = - \frac{ \Delta \D_L }{\bar \D_L}\;. 
\end{equation}
It is worth noticing that the luminosity distance is also related to the signal-to-noise ratio ($\sigma^2$) for the detection of gravitational waves (see e.g.~\cite{Finn:1992wt, Poisson:1995ef}), via
\begin{equation}
\sigma^2 = 4\int^\infty_0 \frac{|\tilde h|^2}{S_n} \,\, d \, f_{r} \, ,
\end{equation}
where $\tilde h=  (\M_r^2/\D_L) (f_r \M_r)^{-7/6} e^{i\psi_r}$ is the Fourier transform of Eq.~(\ref{h2}),  $S_n$ is the spectral noise density and $\psi_r \equiv 2\,\pi\,f_{r}\,t_o + \phi_r(t_o)$ with $t_o$ being the stationary point of the phase~\cite{Laguna:2009re}
\footnote{Following \cite{Laguna:2009re}, we have used the stationary-phase approximation and neglected the antenna-pattern functions.}
.
Hence we have~\cite{Poisson:1995ef} 
\begin{equation}
\sigma^2 = 4\left(\frac{{\cal M}_r}{\D_L}\right)^2\int^\infty_0  \frac{(f_r\, {\cal M}_r)^{-7/3}}{(S_n/{\cal M}_r)}\,d(f_r\, {\cal M}_r) \, ,
\end{equation}
and
\begin{equation}
\label{deltasigma}
\frac{\Delta \sigma}{\bar \sigma} = - \frac{ \Delta \D_L }{\bar \D_L} \; .
\end{equation}

In the rest of the paper we will drop the unobservable constant contribution evaluated at the observer, denoted with a subscript zero.
Finally, it is important to point out that using this prescription one can generalize the definition of $\Q$ of Eq. (\ref{Q}) and $\psi_r$ by adding all Post-Newtonian corrections to the wave amplitude and phase of gravitational waves (see e.g.~\cite{Thorne:1969rba, Turner:1978zz, Kovacs:1978eu, Blanchet:1989fg, Lincoln:1990ji, Blanchet:1992br, Poisson:1995ef}).


\section{Results}
\label{sec:results}
Here we compute the modifications of the value of the luminosity density $\D_L$ inferred from gravitational waves, due to perturbations.


We can make Eq. (\ref{deltaDL}) explicit, i.e. we can write the correction to the luminosity distance as
 \begin{eqnarray}
\label{DL_2}
\frac{ \Delta \D_L}{\bar \D_L} &=& \left(1-\frac{1}{\cH \bar \chi} \right)  v_\| -    \frac{1}{2}  \int_0^{\bar \chi} \ud \tilde \chi  \frac{ \left(\bar \chi-\tilde \chi\right)}{\tilde \chi \bar \chi} \,  \triangle_\Omega \left( \Phi + \Psi \right) + \nonumber\\
&+&\frac{1}{\cH \bar \chi} \Phi - \left(1-\frac{1}{\cH \bar \chi} \right) \int_0^{\bar \chi} \ud\tilde \chi \left(\Psi'+\Phi'\right) -\left(\Phi + \Psi \right) + \frac{1}{\bar \chi} \int_0^{\bar \chi} \ud \tilde \chi \left(\Phi + \Psi \right) \; ,
\end{eqnarray} 
where
$ \triangle_\Omega\equiv\bar \chi^2 \bar \nabla^2_\perp = \bar \chi^2 (\bar \nabla^2
- \bar \partial_\parallel^2 - 2 {\bar \chi}^{-1} \bar \partial_\parallel)= (\cot \partial_\theta + \partial_\theta^2 +\partial_\varphi/\sin^2\theta)$.
We can recognize in Equation~\eqref{DL_2} the presence of a velocity term (the first r.h.s. term), followed by a lensing contribution, and the final four terms account for SW, ISW, volume and Shapiro time-delay effects. \\
To numerically compute the magnitude of this effect, we calculate the mean 
fluctuation of the effect, at any given redshift, as
\begin{equation}
\label{eq:Clf}
C_\ell^{\rm \D_L}=\VEV {\frac{\Delta \D_L}{\bar \D_L} \; \frac{\Delta \D_L}{\bar \D_L}^*}= {2 \over \pi} \int \ud k\, k^2 \left[ I_{\ell}^{\D_L }(k)\right]^2  P_\Psi(k)
 \, ,
\end{equation}
where
\begin{eqnarray}
 I_\ell^{ \D_L}&=&\frac{9}{10} T_m(k)  \int \!\! \ud \bar \chi~ \bar\chi^2\, {\W}_\chi \Bigg\{- j_\ell (\bar \chi k)\left[ \frac{\Gtf_\Psi(\bar a, k)}{\bar a}  +\left(1-\frac{1}{\cH \bar \chi} \right)  \frac{\Gtf_\Phi(\bar a, k)}{\bar a} \right]  \nonumber\\
&& - \left(1-\frac{1}{\cH \bar \chi} \right) \left[\frac{\ell}{\bar \chi k} j_\ell (\bar \chi k)- j_{\ell+1} (\bar \chi k)\right]\Gtf_v(\bar a, k) +   \int_0^{\bar \chi} \ud \tilde \chi\,j_\ell (\tilde \chi k) \Bigg[   \frac{1}{\bar \chi}  \left(\frac{\Gtf_\Psi(\tilde a, k)+\Gtf_\Phi(\tilde a, k)}{\tilde a}\right)\nonumber\\
&&  - \left(1-\frac{1}{\cH \bar \chi} \right) \tilde a \cH(\tilde a) \frac{\ud }{\ud \tilde a}\left(\frac{\Gtf_\Psi(\tilde a, k)+\Gtf_\Phi(\tilde a, k)}{\tilde a}\right) + \frac{\ell (\ell+1)\left(\chi-\tilde \chi\right)}{2 \chi \tilde \chi}\left(\frac{\Gtf_\Psi(\tilde a, k)+\Gtf_\Phi(\tilde a, k)}{\tilde a}\right) \Bigg] \Bigg\}\;.\nonumber\\
\end{eqnarray}

Here $\W_\chi$ represents the normalized object selection function (the normalization convention is $\int \ud \chi \, \chi^2 \, {\W}_\chi=1$)\footnote{${\W}_\chi$ can be easily related to the redshift distribution ${\W}_z = \chi^2 {\W}_\chi/a \cH$, where $\int \ud z \, {\W}_z=1$.
}, $j_\ell(x)$ are spherical Bessel functions of order $\ell$ and argument $x$ and the quantities $\Gtf_\Phi$, $\Gtf_\Psi$ $\Gtf_m$ and $\Gtf_v $ are defined in Appendix \ref{PowerSpectra}.

In Figure~\ref{fig:Cl_terms} we show the relative importance of different contributions to Eq.~\eqref{DL_2}: velocity (dotted lines and square symbols), lensing dashed lines and circles), and gravitational potentials (as the sum of SW, ISW, volume and Shapiro time-delay effects; dot-dashed lines and diamonds).
We plot results computed by calculating the $C_\ell$ of Eq.~\eqref{eq:Clf} for sources on a variety of redshift ranges, $z\in\{[0,1],[1,2],[3,4],[4,5] \}$; in the plots, colors indicate different values of $z$ within the range (in ascending order, black, red, blue, green, purple).

We can see how the velocity terms are the most important on large scales and low redshift (in analogy with the velocity contributions to  galaxy clustering, see~\cite{Raccanelli:2016Doppler}).
The gravitational potentials contributions are always very small; they become more important than velocity ones at higher-$z$, but they stay almost two orders of magnitude below the most important contribution at each redshifts and scales, at maximum.
Lensing terms are in general the dominant ones; being an integrated (along the line of sight) quantity, their importance is lowest al lowest $z$, but they quickly overpower velocity terms (for $\ell>20$, $z>0.5$).

\begin{figure*}
\centering
\includegraphics[width=0.47\columnwidth]{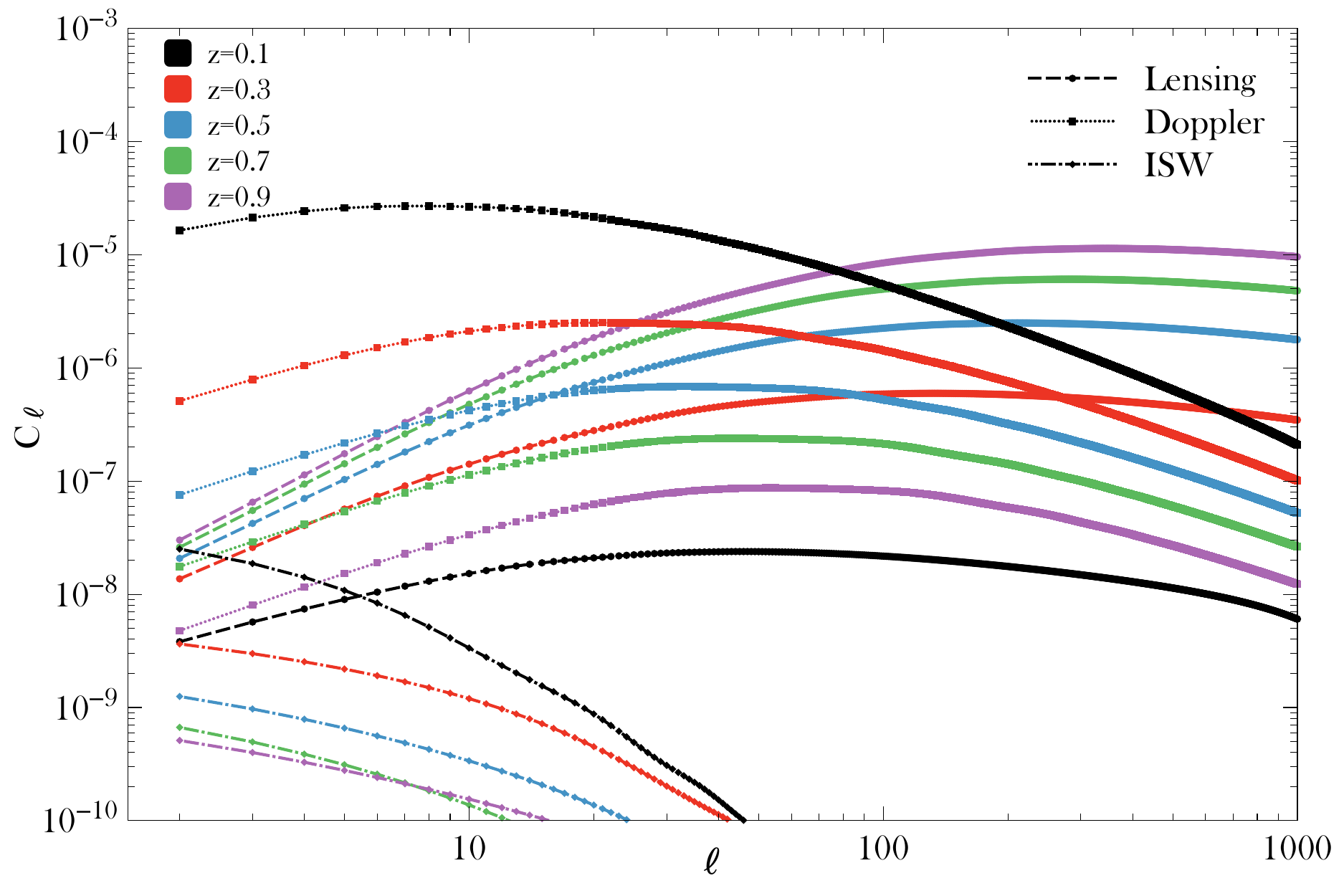}
\includegraphics[width=0.47\columnwidth]{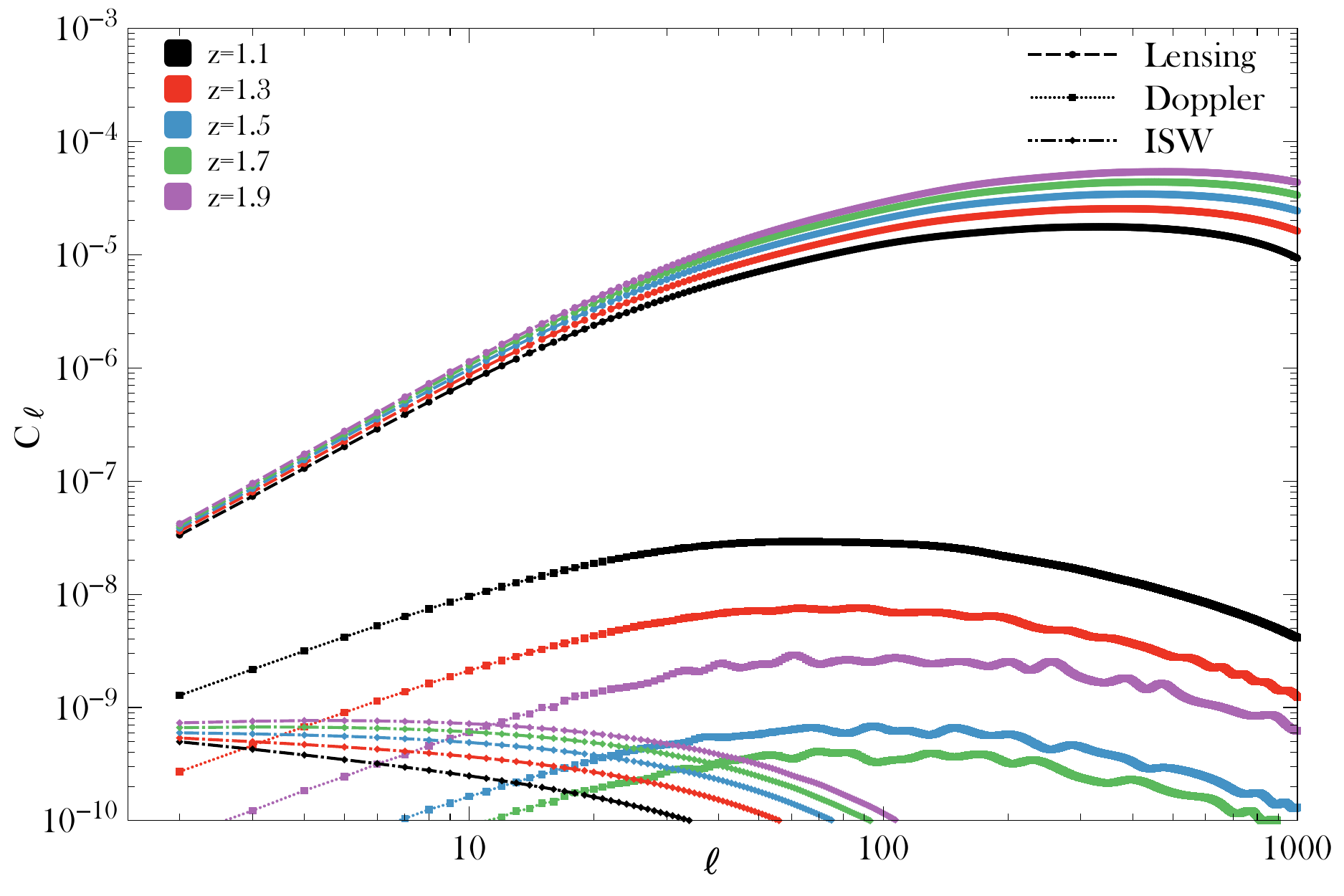}
\includegraphics[width=0.47\columnwidth]{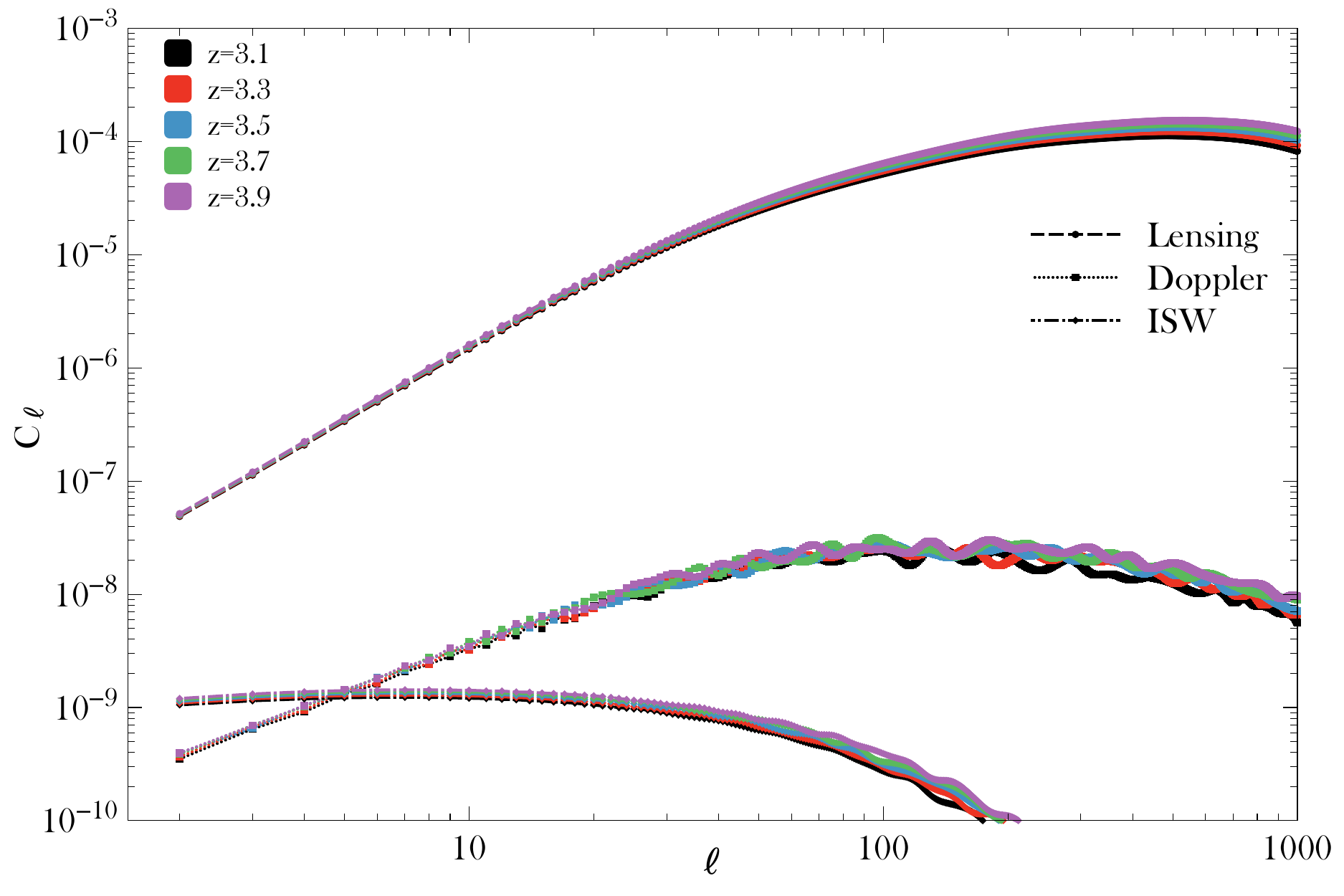}
\includegraphics[width=0.47\columnwidth]{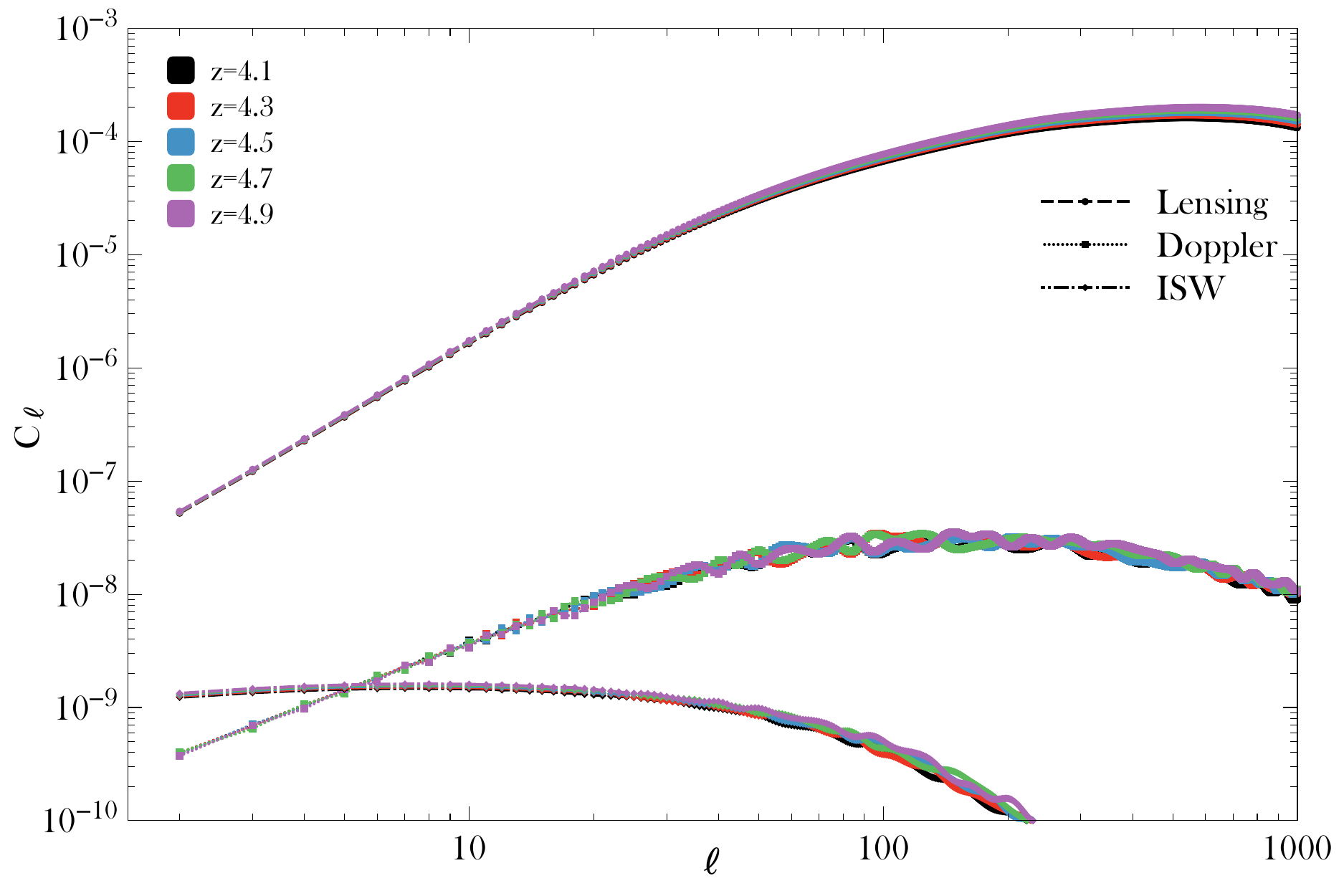}
\caption{Different contributions to the modification of $\D_L$, for a variety of redshift ranges. For the expression of different terms, see Eq.~\eqref{DL_2}.}  
\label{fig:Cl_terms}
\end{figure*}

\begin{figure*}
\centering
\includegraphics[width=0.95\columnwidth]{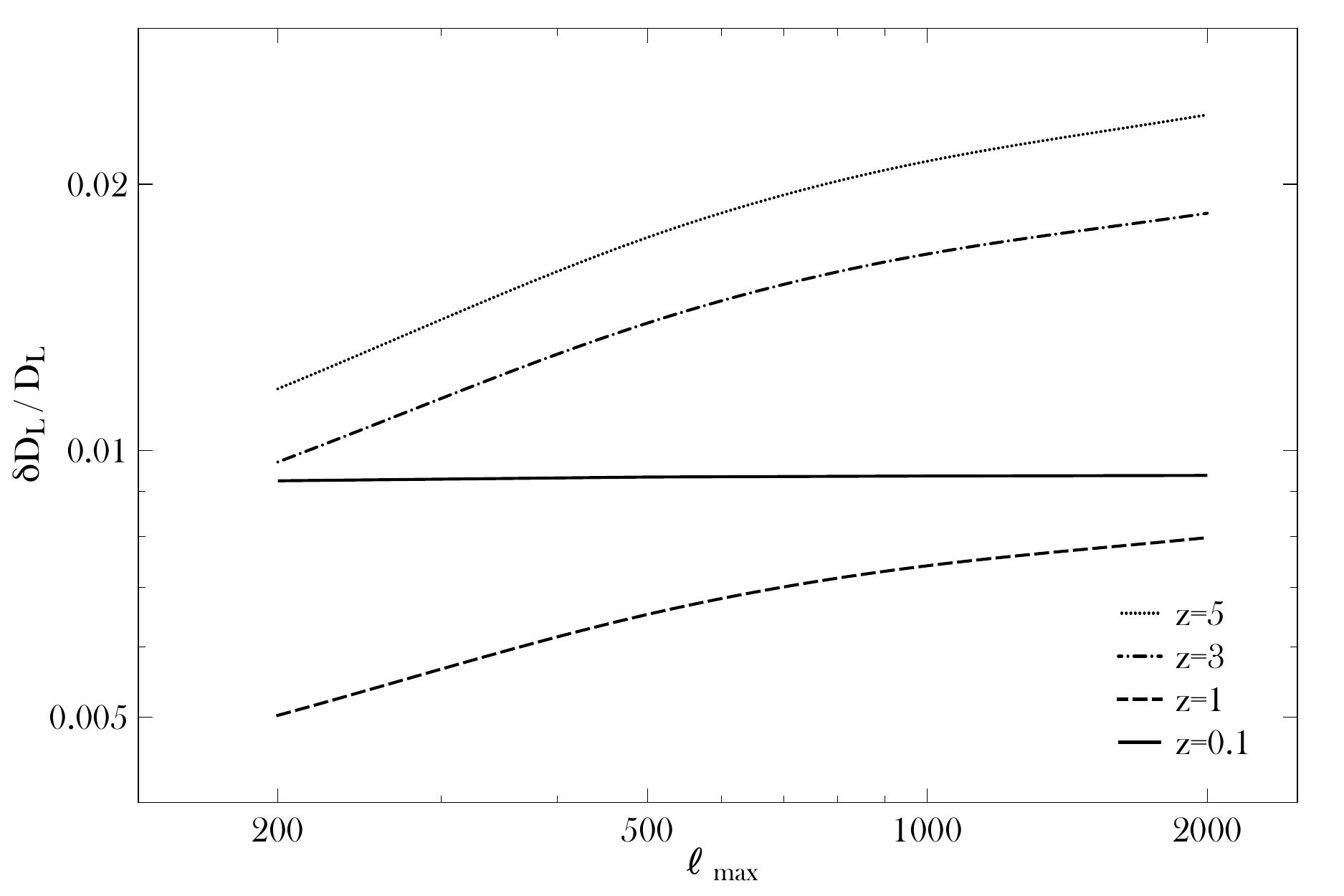}
\caption{Total correction to the luminosity distance as a function of the maximum $\ell$ used, for different redshifts.}  
\label{fig:DL_ellmax}
\end{figure*}

It can be seen from Figure~\ref{fig:Cl_terms} that different contributions have different behavior at different scales; therefore, given that the total result is computed as a sum over $\ell$, in Figure~\ref{fig:DL_ellmax} we show how the luminosity distance correction depends on the maximum multipole used.
The plot shows that the results are only mildly dependent on the scale selected; in the following we will assume $\ell_{\rm max}=1000$.
Taking into account corrections arising in the non-linear regime, lensing corrections are larger (for photons the corrections are estimated to be $\sim0.05 z$ magnitudes, see e.g.~\cite{Quartin:2013moa, BenDayan:2013gc}), given that non-linear lensing contributions are dominant over linear perturbations~\cite{Tamanini:2016zlh}. However, in the non-linear regime, there are large effects on the luminosity distance-redshift relation and on the chirp mass estimate~\cite{Bonvin:2016qxr}, due to local peculiar velocities of the merging binary systems. For this reason we limit our analysis to large, linear, scales. 

The final result is then computed by calculating:
\begin{equation}
\label{eq:rmsf}
\daleth_\lamed = \left(\nsum_{\ell} \frac{2\ell+1}{4\pi}C_\ell^{\rm DL}\right)^{1/2} \, .
\end{equation}

We evaluate the $\daleth_\lamed$ contribution for a variety of future GW experiments, such as the Einstein Telescope (ET), DECIGO and the Big-Bang Observer.

\begin{figure*}
\centering
\includegraphics[width=0.95\columnwidth]{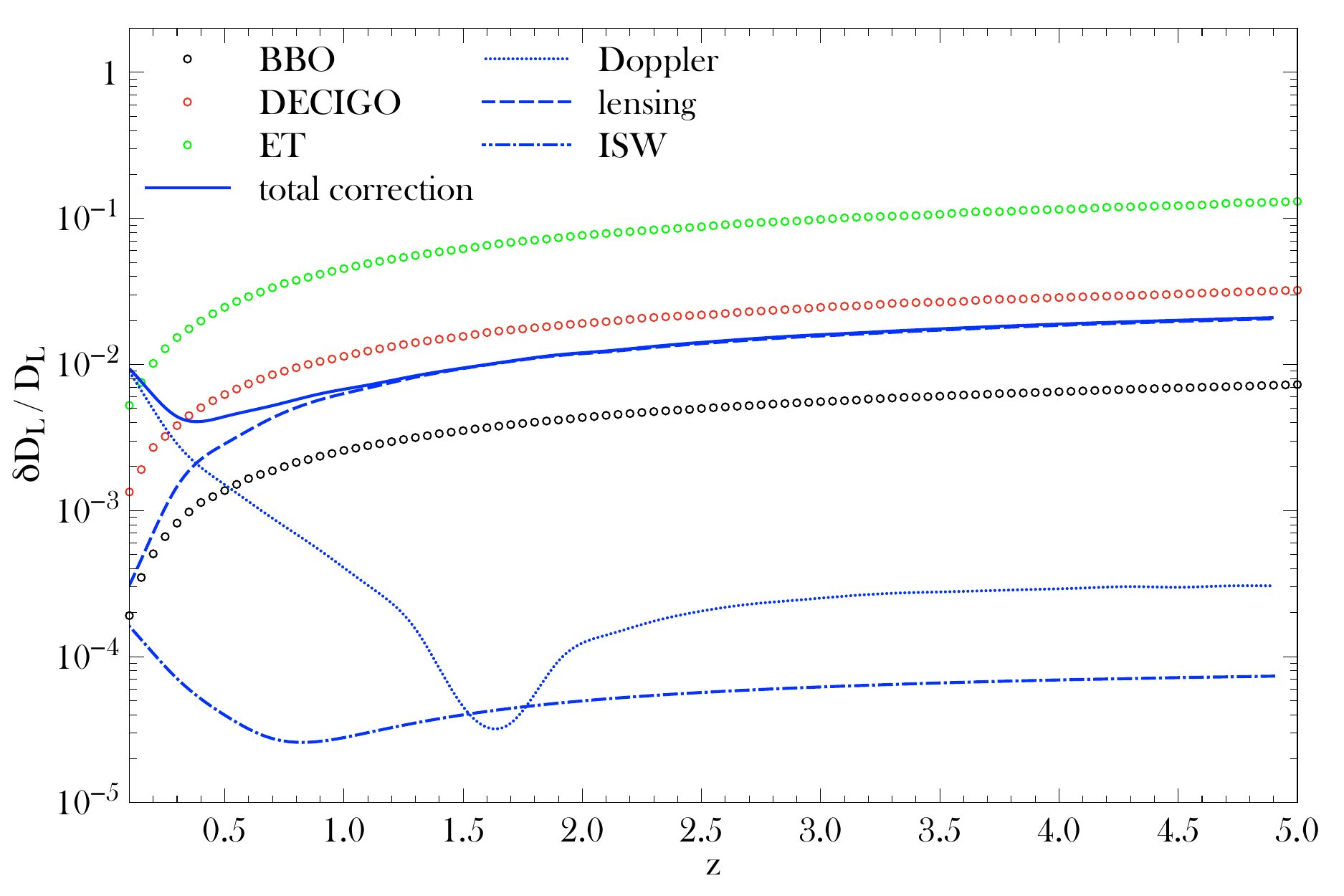}
\caption{Total correction to luminosity distance estimates due to perturbations (computed as in Eq.~\eqref{eq:rmsf}), as a function of $z$. The dotted, dashed and dot-dashed lines show Doppler, lensing and ISW-like contributions, respectively, while the solid line shows the total effect.
Points show the predicted precision in measurements of the luminosity distance, at any redshift, for the Einstein Telescope (green points), DECIGO (red points) and the Big Bang Observer (black points).
}  
\label{fig:final}
\end{figure*}

In Figure~\ref{fig:final} we show the total correction to luminosity distance estimates due to perturbations, as a function of $z$. The dotted, dashed and dot-dashed lines show velocity, lensing and ISW-like contributions, respectively, while the solid line shows the total effect.
Points show the predicted precision in measurements of the luminosity distance, at any redshift, for the Einstein Telescope (green points), DECIGO (red points) and the Big Bang Observer (black points), taken from~\cite{Cutler:2009qv, Camera:2013xfa}.

We can see that the additional $\D_L$ uncertainty due to the inclusion of perturbations has a peak at low-$z$ due to velocity contributions, that surpasses the predicted measurement errors for all the experiments considered here. However, velocity effects rapidly decrease and lensing takes over.
The integrated lensing effects increase with $z$, and their amplitude is of a factor$\sim2$ smaller than DECIGO forecast precision.
When we consider BBO, however, the correction to the $\D_L$ uncertainty is consistently twice the predicted errors, making it a very relevant correction, that one will need to take into account.


\section{Conclusions}
\label{sec:conclusions}
In this paper we investigated the effect of including perturbations in the estimate of luminosity distances as inferred from gravitational wave observations.
While the usual calculation is performed in a homogeneous and isotropic FLRW Universe, we show that the {\it observed} GWs have a different luminosity distance $\D_L$ than the {\it real-space} ones, in the same way {\it observed (redshift)}-space galaxies are in a different location than the ones in real space.
We derive expressions for the luminosity distance in what we call {\it Redshift-GW} frame and the difference between that and the unperturbed case, in analogy with what has been done for photons (see e.g.~\cite{Bonvin:2005ps, Fleury:2016fda}), using the Cosmic Rulers formalism to generalize the results of~\cite{Laguna:2009re} by including all velocity, lensing and gravitational potentials (Sachs-Wolfe, Integrated Sachs-Wolfe, volume distortion and Shapiro time-delay effects).

The main difference w.r.t. \cite{Laguna:2009re} is that the amplitude of the GW at the initial condition is not evaluated {\it in background} frame, but in  {\it Redshift-GW} frame.
In this way  we are able to connect our results with the initial amplitude, using directly the output from simulations of coalescing binary BHs, which produce the templates.

The inclusion of deviations due to perturbations therefore causes an additional uncertainty in the determination of $\D_L$ from GW observations; we then computed the root-mean square fluctuations of this effect for a wide range of redshifts.

Our results show that, as expected, the dominant source of correction is due to lensing magnification effects, again in analogy with the case of photons and galaxy number counts (see e.g.~\cite{Raccanelli:2015vla}).
The amplitude of this effect is however small, and it will not constitute a relevant additional uncertainty for near future and next generation GW experiments.
However, for future GW detectors such as the planned Big Bang Observer, the uncertainty we compute is predicted to be as much as twice the normally predicted error in $\D_L$, making it necessary to be included in any realistic analysis.
Finally, the inclusion of perturbations and the distinction between real space and Redshift-GW frame might affect other aspects of GW astronomy; we leave this investigation to a future work.

\vspace{0.5 cm}

\noindent{\bf Acknowledgments:}\\
We thank Ilias Cholis for useful suggestions during the development of this project, Donghui Jeong, Julian Mu\~{n}oz and Helvi Witek for interesting discussions. \\
During the preparation of this work DB was supported by the Deutsche Forschungsgemeinschaft through the Transregio 33, The Dark Universe. \\
AR has received funding from the People Programme (Marie Curie Actions) of the European Union H2020 Programme under REA grant agreement number 706896 (COSMOFLAGS). Funding for this work was partially provided by the Spanish MINECO under MDM-2014-0369 of ICCUB (Unidad de Excelencia ``Maria de Maeztu'') and the Templeton Foundation.

\appendix

\section{Perturbation terms in the Poisson Gauge}\label{Poiss-pert}

From Eq.\ (\ref{Poiss-metric}), the perturbations of  $\tilde g_{\mu \nu}$, $\tilde g^{\mu \nu}$ and   the comoving metric  $\hat g_{\mu \nu}$, $\hat g^{\mu \nu}$ are
\begin{eqnarray}
\begin{array} {lll}
\tilde g_{00}= a^2 \hat g_{00}=- a^2 \left(1+ 2 \Phi \right),  & \quad& \tilde g^{00}=a^{-2} \hat g^{00}=- a^{-2} \left(1-2  \Phi \right), \\ \\
\tilde g_{0i}=a^2 \hat g_{0i}= 0\;,  & \quad&  \tilde g^{0i}=a^{-2} \hat g^{0i}=0, \\ \\
\tilde g_{ij}=a^2 \hat g_{ij}= a^2 \left(\delta_{ij} -2 \delta_{ij} \Psi  \right),  & \quad&  \tilde g^{ij}=a^{-2}\hat g^{ij}=a^{-2}  \left( \delta^{ij}+2 \delta^{ij} \Psi \right). \\
  \end{array} 
\end{eqnarray}

For Christoffel symbols $\tilde \Gamma^\mu_{\rho \sigma}=\tilde \Gamma^{\mu (0)}_{\rho \sigma}+\tilde \Gamma^{\mu (1)}_{\rho \sigma}$ (using   $\tilde g_{\mu \nu}$)  and $\hat \Gamma^\mu_{\rho \sigma}=\hat \Gamma^{\mu (0)}_{\rho \sigma}+\hat \Gamma^{\mu (1)}_{\rho \sigma}$ (using $\hat g_{\mu \nu}$), we obtain

\begin{eqnarray}
\begin{array} {lll} 
\tilde \Gamma^{0 (0)}_{00}=\cH\,,   \quad   \quad  &\tilde  \Gamma^{0 (0)}_{0i}=0\,,   \quad   \quad  &\tilde \Gamma^{i (0)}_{00}=0\,, \\ \\
\tilde \Gamma^{i (0)}_{00}=0\,,   \quad   \quad  & \tilde \Gamma^{i (0)}_{j0}=\cH \delta^i_j\,,  \quad    \quad &\tilde \Gamma^{i (0)}_{jk}=0\,, 
~~~~ \hat \Gamma^{\mu (0)}_{\rho \sigma} = 0\,,
\end{array}  
\end{eqnarray}

\begin{eqnarray}
\begin{array} {lll} 
\tilde\Gamma^{0 (1)}_{00}=\hat \Gamma^{0 (1)}_{00}  &  \quad \quad   & \tilde \Gamma^{0 (1)}_{0i}=\hat \Gamma^{0 (1)}_{0i}  \\ \\
\tilde \Gamma^{0 (1)}_{ij}=\hat \Gamma^{0 (1)}_{ij} - 2 \cH \left( \Phi  + \Psi  \right) \delta_{ij}&  \quad \quad  &\tilde \Gamma^{i (1)}_{00}=\hat \tilde\Gamma^{i (1)}_{00}\,  \\ \\
\tilde \Gamma^{i (1)}_{j0}=\hat \Gamma^{i (1)}_{j0}\,  &  \quad  \quad & \tilde  \Gamma^{i (1)}_{jk}=\hat \Gamma^{i (1)}_{jk} \,, \\ \\
\hat \Gamma^{0 (1)}_{00} = \Phi' \,,&  \quad  \quad & \hat \Gamma^{0 (1)}_{0i} = \p_i \Phi \,,\\ \\
\hat \Gamma^{0 (1)}_{ij}  =- \delta_{ij} \Psi' \,,&  \quad  \quad & \hat \Gamma^{i (1)}_{00} = \p^i \Phi \,, \\ \\
\hat \Gamma^{i (1)}_{j0} = -  \delta^i_{j} \Psi'\,, &  \quad  \quad & \hat \Gamma^{i (1)}_{jk} = -  \delta_k^i \p_j \Psi-  \delta_j^i\p_k\Psi + \delta_{jk} \p^i \Psi  \,,
\end{array}  
\end{eqnarray}

For four-velocity $u^\mu$, we find
\begin{eqnarray}
\label{Poiss-u0i}
u_0&=&-a\left(1+\Phi\right) ,   \quad   \quad u_i=a v_i   \quad   \quad u^0=\frac{1}{a}\left(1-\Phi \right) ,  \quad   \quad u^i =\frac{1}{a}v^{i}\;.
\end{eqnarray}

For the tetrad:

\begin{eqnarray} \label{Poiss-LambdaE-1-2}
 E_{\hat 0 0}^{(1)}= - \Phi \;,    \quad   \quad  E_{\hat 0 i}^{(1)}=  v_i \;,     \quad   \quad
 E_{\hat a 0}^{(1)}= - v_{\hat a} \;,    \quad   \quad  E_{\hat a i}^{(1)}=- \delta_{\hat a i}\Psi \;,   
\end{eqnarray}

\section{Power Spectra relations} \label{PowerSpectra}

In order to encode all possible DE models let us define in Fourier space the following relations
\begin{align}
v(a,\bf{k}) &= - \frac{9}{10} \frac{T_m(k)}{k} \Gtf_v (a,k) \Psi_p(\bf{k}), \\
\delta^{C}_m(a,\bf{k}) &= - \frac{9}{10} T_m(k) \Gtf_m (a,k) \Psi_p(\bf{k}), \\
\Psi(a,\bf{k}) &=\frac{9}{10} T_m(k) \frac{\Gtf_\Psi (a,k)}{a} \Psi_p(\bf{k}), \\
\Phi(a,\bf{k}) &=\frac{9}{10} T_m(k) \frac{\Gtf_\Phi (a,k)}{a} \Psi_p(\bf{k}),
\end{align}
where $\Gtf$ are suitable transfer functions and depend on the model that we are considering and $T_m(k)$ is the Eisenstein Hu transfer function \cite{Eisenstein:1997jh} (or BBKS \cite{Bardeen:1985tr}).

Here $\Psi_p(\bf{k}) $ is the primordial value set during inflation, whose power spectrum
\be
 \left\langle\Psi_p^* ({\bf k}_1)\Psi_p ({\bf k}_2) \right\rangle=(2 \pi)^3\delta_D^3({\bf k}_1-{\bf k}_2) P_\Psi(k_1).
 \ee
reads
 \be
P_\Psi(k)= \frac{50}{9} \pi^2 A^2 \delta_H \left[\frac{\Omega_m(a=1)}{\Gtf_\Psi(a=1,k)}\right]^2 k^{-3} \left(\frac{k}{\ho}\right)^{n-1} \, .
 \ee

 $\Gtf_\Phi$, $\Gtf_\Psi$ $\Gtf_m$ and $\Gtf_v $ are, in principle, all functions of space and time. For $\Lambda$CDM and Dark energy + Dark matter models, we can write them as
\begin{align}
   \Gtf_\Phi&=\Gtf_\Psi = D_m \label{eq:Gphi}\\
     \Gtf_m&=\frac23  \frac{k^2}{\Omo  H_0^2} D_m\, \\
\Gtf_v&=f \frac{\cH}{k}\Gtf_m.\label{eq:Gv}
\end{align}
where 
$D_m (a)$ is the growth function \citep[defined as in ][]{Dodelson:2003ft} and\footnote{In particular, $D_m$ is normalised as $D_m(a_{\rm DM})=a_{\rm DM}$, where $a_{\rm DM} =a (\tau_{\rm DM})$ and $\Omega_{\rm DM}=1$.}
\be
f_m(a)=\frac{\ud \ln\delta_m^{C}}{\ud \ln a}=\frac{\ud \ln \Gtf_m}{\ud \ln a}= \frac{\ud \ln D_m}{\ud \ln a}
\ee
is usually referred to as the growth factor.
Here at background level,
 \begin{align}
 \cH^2&=a^2 \cH_0^2 \left(\Omo a^{-3}+\OLo\right), \\
 \frac{\cH'}{\cH^2}&=\left(1-\frac{3}{2} \Omo a^{-3}\right),
\end{align}
and, at the first perturbative order, we have
 \begin{align}
 {\delta^P_m}'+\nabla^2 v -3\Phi' &=0, \\
 v'+\cH v+\Psi &=0,\\
\nabla^2 \Phi &=4\pi G \bar \rho_m \delta_m^{C},\label{eq:poisson}\\
\Psi&=\Phi,\label{eq:slip}
\end{align}

where we have defined  $\delta_m^P$ as the matter overdensity in Poisson gauge and
\be
\delta_m^{C}=\delta_m^P-3\cH v
\ee
 the gauge-invariant comoving density contrast.

\section{Dependence on maximum multipole}
\label{sec:ellmax}
\begin{figure*}
\centering
\includegraphics[width=0.95\columnwidth]{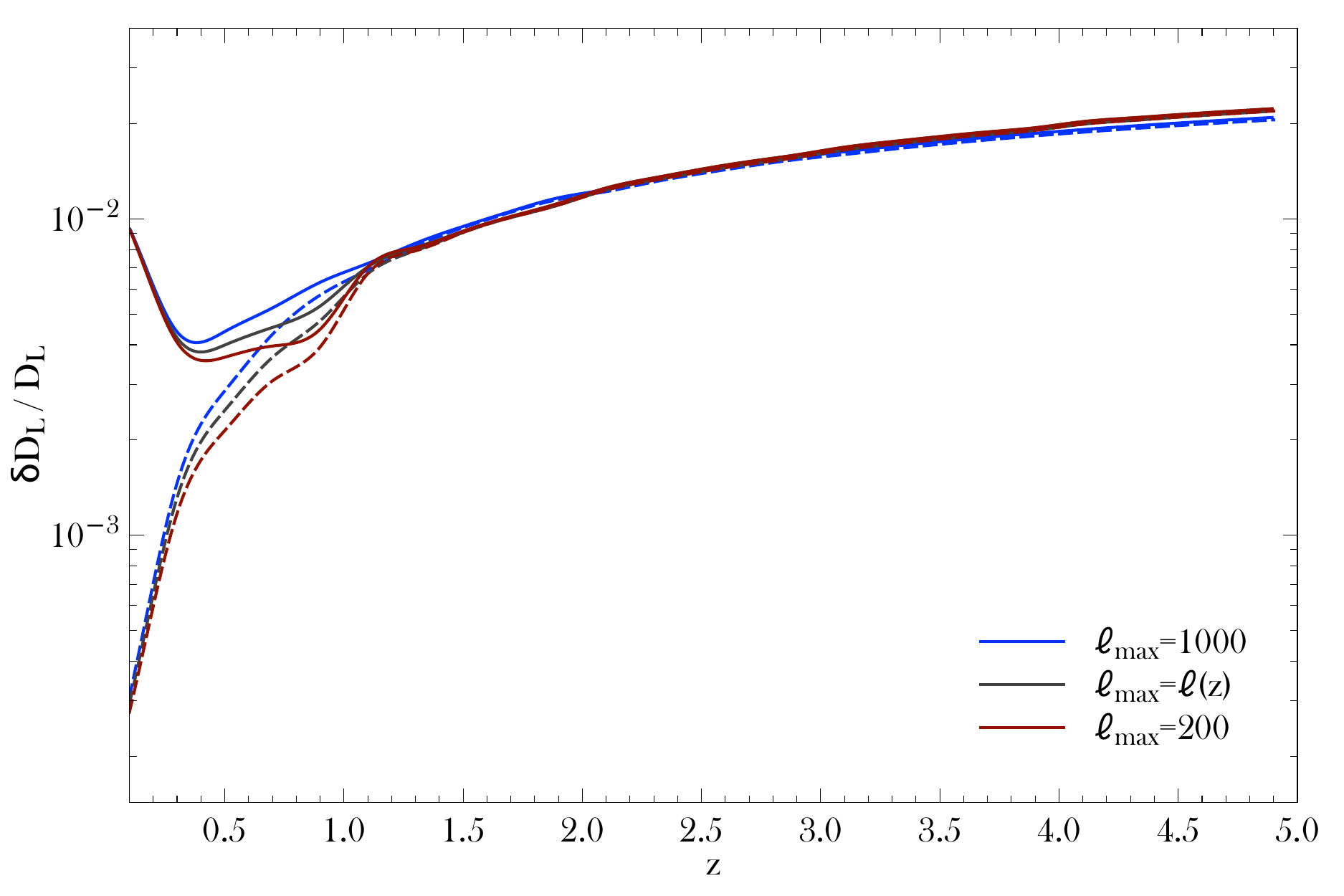}
\caption{Total correction to luminosity distance estimates due to perturbations, as a function of $z$, for three different cases of $\ell_{\rm max}$.
Solid lines show the total effect, while the dashed one refers to the lensing contributions. Doppler and ISW effects are subdominant and they are important mostly at very large scales, hence not affected by the choice of the smallest scale used.}  
\label{fig:ellmax}
\end{figure*}

Here we show how the results change as a function of the maximum multipole $\ell_{\rm max}$ used in our calculations.
In Figure~\ref{fig:ellmax} we show the results, for the lensing component and the total contributions, in the fiducial case of $\ell_{\rm max}=1000$ used in the main text, along with the $\ell_{\rm max}=200$, and a $\ell_{\rm max}=\ell_{\rm max}(z)$ cases. In the latter, we assume $\ell_{\rm max}=200, 800, 1100, 1300, 1500$, for the different bins.

As it can be seen, while the lensing can change by a factor or $\sim2$ in some cases, this happens only in a limited range of (mostly relatively low) redshifts, and do not significantly change the results of the paper.
Doppler and ISW contributions are not shown as they are subdominant and important mostly at very large scales, hence not affected by the choice of the smallest scale used.
Overall it is evident that the corrections will be larger than the statistical errors for the BBO experiment, even limiting to very large scales.

\end{document}